\def\year{2022}\relax

\documentclass[letterpaper]{article} 
\usepackage{aaai22}  
\usepackage{times}  
\usepackage{helvet}  
\usepackage{courier}  
\usepackage[hyphens]{url}  
\usepackage{graphicx} 
\usepackage{natbib}  
\usepackage{caption} 

\usepackage{subfigure}

\usepackage{verbatim}
\usepackage{multirow}
\usepackage{makecell}
\usepackage{url}
\usepackage{xcolor}

\DeclareCaptionStyle{ruled}{labelfont=normalfont,labelsep=colon,strut=off} 
\usepackage{algorithm}
\usepackage{algorithmic}

\makeatletter
\newcommand{\printfnsymbol}[1]{%
  \textsuperscript{\@fnsymbol{#1}}%
}
\makeatother

\usepackage{newfloat}
\usepackage{listings}
\floatstyle{ruled}
\newfloat{listing}{tb}{lst}{}
\floatname{listing}{Listing}

\newcommand{\czl}[1]{{\textcolor{red}{ [#1]}}}

\title{Shorter Is Different: Characterizing the Dynamics of \\ Short-Form Video Platforms}

\author{
    Zhilong Chen\equalcontrib, Peijie Liu\equalcontrib, Jinghua Piao, Fengli Xu\thanks{Corresponding authors.}, Yong Li\printfnsymbol{2}
}
\affiliations{

    Department of Electronic Engineering, BNRist, Tsinghua University, China\\
    fenglixu@tsinghua.edu.cn, liyong07@tsinghua.edu.cn
}

\begin{document}

\maketitle

\begin{abstract}
The emerging short-form video platforms have been growing tremendously and become one of the leading social media recently. Although the expanded popularity of these platforms has attracted increasing research attention, there has been a lack of understanding of whether and how they deviate from traditional long-form video-sharing platforms such as YouTube and Bilibili. To address this, we conduct a large-scale data-driven analysis of Kuaishou, one of the largest short-form video platforms in China. Based on 248 million videos uploaded to the platform across all categories, we identify their notable differences from long-form video platforms through a comparison study with Bilibili, a leading long-form video platform in China. We find that videos are shortened by multiples on Kuaishou, with distinctive categorical distributions over-represented by life-related rather than interest-based videos. Users interact with videos less per view, but top videos can even more effectively acquire users' collective attention. More importantly, ordinary content creators have higher probabilities of producing hit videos. Our results shed light on the uniqueness of short-form video platforms and pave the way for future research and design for better short-form video ecology.
\end{abstract}

\section{Introduction}

Short-form video platforms have been growing at a phenomenal speed and become increasingly important in the social media ecology in recent years. TikTok, one of the most important and representative short-form video platforms across the globe, has acquired a user base of more than 1.5 billion by 2023 with a 16\% year-wise growth~\cite{2024tiktok}. In China alone, the daily active users of Kuaishou (the Chinese version of Kwai) have reached 300 million by 2020~\cite{2020kuaishou}. Through videos ranging from seconds to minutes, users can actively express their creativity and joy and relatively easily share their lives and talents~\cite{herrman2019tiktok}. 

The impressive boom and accumulating prominence of short-form videos have stimulated the recent surge of related studies. Researchers have been dedicated to unveil user motivations~\cite{lu2019fifteen}, uses and non-uses~\cite{lu2020exploring}, and user conceptions of the recommendation algorithms~\cite{barta2021constructing,lee2022algorithmic,karizat2021algorithmic}, test user assumptions about trending~\cite{klug2021trick}, and probe into factors influencing virality~\cite{ling2022slapping}. However, these existing researches have not addressed a fundamental question: whether and how short-form video platforms deviate from video-sharing platforms with typical length (which we refer to as long-form video platforms in the following). In other words, \textbf{what makes short-form video excel from traditional video-sharing platforms, i.e., what are the defining characteristics of short-form video platforms?} 

This work seeks to fill this gap by measuring the core dimensions of short-form video platforms. Specifically, for the videos themselves, \textbf{what are the videos like on short-form platforms?} (RQ1) When the interactions from the user side are taken into account, \textbf{what are the patterns for user engagement with these videos?} (RQ2) When the competitions between videos are further considered and a more holistic ecology-based view is taken, \textbf{how is collective attention allocated across these videos?} (RQ3) Most importantly, across these dimensions, \textbf{how do short-form video platforms resemble and vary from traditional long-form video platforms}?

To address the aforementioned aspects of concern, we conduct a large-scale data-driven analysis of Kuaishou, one of the largest short-form video platforms in China. Building upon data on 248 million short-form videos on Kuaishou and comparing it with data of 4.7 million videos on Bilibili, a leading long-form video platform in China, we find evident discrepancies between the two platforms across all the explored dimensions. 

In terms of the basic characteristics of the videos, we find that the videos on Kuaishou are several times shorter than those on Bilibili. The categorical distributions of Kuaishou and Bilibili videos are significantly different, where Kuaishou is relatively over-represented by life-related videos rather than interest-based ones such as gaming. In terms of user interactions, the two platforms share largely similar overall patterns for the relationship between video lengths and likes and sharing, but the inflection points are different, and like rates and share rates are larger on Kuaishou, which indicates higher levels of explicit engagement per view. However, in terms of the allocation of collective attention, the video-level distribution of view counts is more uneven, where top videos can effectively acquire more views. On the creator level, view counts of videos from the same creator vary more significantly on Kuaishou, leaving a relatively higher probability for ordinary content creators to produce hit videos. These results unveil the fundamental characteristics of short-form video platforms and highlight their distinctiveness from the traditional long-form video-sharing platforms.

We organize the rest of the paper as follows. We start by reviewing related literature. Then, we introduce the background information about Kuaishou and the basics of the datasets we use for our study. Next, we delve into the basic video characteristics, user engagement, and collective attention allocation patterns on Kuaishou and demonstrate its similarities with and discrepancies from Bilibili. Finally, we discuss the distinctiveness of short-form videos and highlight possible implications from our empirical analyses.
\section{Related Work}
In this section, we review three bodies of studies most relevant to the current study: measurements of social media, the ecology of video-sharing platforms, and short-form videos.

\subsection{Measurements of Social Media}
Measurement-based analyses are powerful in unveiling the fundamental attributes of social media and have been widely adopted in related research. \citet{cha2007tube} revealed the unique patterns concerning content production, user participation, popularity distribution, and popularity evolution among other features of YouTube and Daum to characterize user-generated content (UGC) Video-on-Demand (VoD) systems. \citet{kwak2010twitter} explored the network topology and practices of information sharing on Twitter to understand its basic mechanisms~\cite{kwak2010twitter}, whereas Cha et al.'s delineation of user influence and Mislove et al.'s depiction of user demographics on Twitter help with better definition of Twitter~\cite{cha2010measuring,mislove2011understanding}. \citet{hu2014we} identified the 8 major types of photo contents and 5 types of users on Instagram, providing a first delineation of the Instagram ecology. Similar research attempts have also been adopted for the understanding of Flickr~\cite{mislove2007measurement}, Microsoft Messenger~\cite{leskovec2008planetary}, location-based services~\cite{noulas2011empirical,xu2018understanding}, Pinterest~\cite{gilbert2013need}, Linkedin~\cite{anderson2015global}, and social commerce platforms~\cite{xu2019think,cao2020your,chen2020understanding,xu2021understanding}, etc. 

Building upon and extending these works, our study aims to borrow the power of measurement-based analyses to quantify and understand short-form video platforms.

\subsection{Ecology of Video-sharing Platforms}
Previous literature has been dedicated to the understanding of the ecology of video-sharing platforms. One central focus of the related studies is video popularity. It has been revealed that the popularity of these platforms follows power laws with truncated tails~\cite{cha2007tube} and early and past views of a video can predict its future popularity~\cite{cha2007tube,szabo2010predicting,pinto2013using}. This popularity shows high geographical locality~\cite{brodersen2012youtube} and is deeply influenced by mechanisms of both learning and network effects~\cite{qiu2015two}. When the evolution patterns of popularity are further considered, existing studies reveal that views of copyright-protected videos concentrate on their early lives, whereas sudden bursts of popularity are more likely to be spotted among top videos~\cite{figueiredo2011tube}. Endogenous relaxation~\cite{crane2008robust} and four-phase based trending transition~\cite{yu2015lifecyle} models have been proposed to explain the dynamics of video popularity. 

With the growing recognition of the essence of content satisfaction in user experiences~\cite{2012youtube}, some scholars have gone beyond popularity to attend to user interaction and engagement with these online videos. \citet{chatzopoulou2010first} indicated the positive relationships among the numbers of views, comments, ratings, and favorites. Similarly, \citet{park2016data} measured user engagement with average watch time and revealed its positive correlation with view counts, like rates, and negative comment sentiment. \citet{wu2018beyond} utilized the duration-normalized average watch percentage of a video to depict the engagement of a video and found it highly predictable and stable over time. 

Some recent endeavors have delved into more specific and nuanced aspects of the video-sharing ecology, characterizing the fake engagement induced by view fraud~\cite{kuchhal2022view}, monetization~\cite{hua2022characterizing}, and implementation of content moderation~\cite{jiang2019bias}, identified how political and controversial content~\cite{hosseinmardi2021examining,lee2022whose}, science~\cite{welbourne2016science,yang2022science,zhang2023understanding}, and even misinformation~\cite{hussein2020measuring,papadamou2022just,micallef2022cross} are delivered and consumed through video-sharing platforms.

Nevertheless, these depictions of the video-sharing ecology are based on the typical long-form video platforms such as YouTube. It remains largely unclear to what extent they apply to the recently emerging short-form video circumstances. Our study intends to fill this gap.

\subsection{Short-form Videos}
With short-form videos becoming increasingly popular and prevalent, research into and based on the context of short-form videos flourishes. Prior research endeavors have been dedicated to users' motivations and categories of engagement~\cite{lu2019fifteen}, uses and non-uses~\cite{lu2020exploring}, and the distribution of posted and popular contents~\cite{shutsko2020user}, offering initial peaks into experiences under short-form videos. Taking the influence of the algorithmic recommendations from ``For You Page'' into account, another line of researchers dig into how the mediation of algorithms influences user experiences on short-form videos. These studies investigated the interplay between the construction of authenticity and algorithmic recommendations~\cite{barta2021constructing}, between algorithmic personalization and users' concepts of the self~\cite{lee2022algorithmic}, and between algorithmic folk theories and identities~\cite{karizat2021algorithmic}, highlighting the challenges facing minority groups such as black~\cite{harris2023honestly} and LGBTQ+ people~\cite{simpson2021you,simpson2022tame,devito2022transfeminine}. However, these studies are limited in scale and adopt primarily qualitative approaches. Although they provide detailed and deep insights into users' encounters with short-form videos, they fall short in generalizability. 

With data at scale, \citet{klug2021trick} quantitatively examined the qualitatively-collected user assumptions about the TikTok algorithm. Based upon videos from the trending section, they confirmed the positive relationship between user engagement and trending as well as upload time and trending, yet common assumptions on the benefits of hashtag usage do not hold. \citet{ling2022slapping} investigated how video content, lifespan, time, and creator profile predict the virality of TikTok videos, and creators' follower numbers, close-up, and medium-shot scale are identified as signs of virality. However, the existing literature fails to unravel the fundamental question of whether and how these short-form videos deviate from long-form videos, which the current work seeks to contribute. 
\section{Background and Data}


\subsection{Background about Kuaishou}
\begin{figure} [htb]
\centering
\subfigure[Examplar landing page of Kuaishou] {
\label{fig:example_ks}
\includegraphics[width=.15\textwidth]{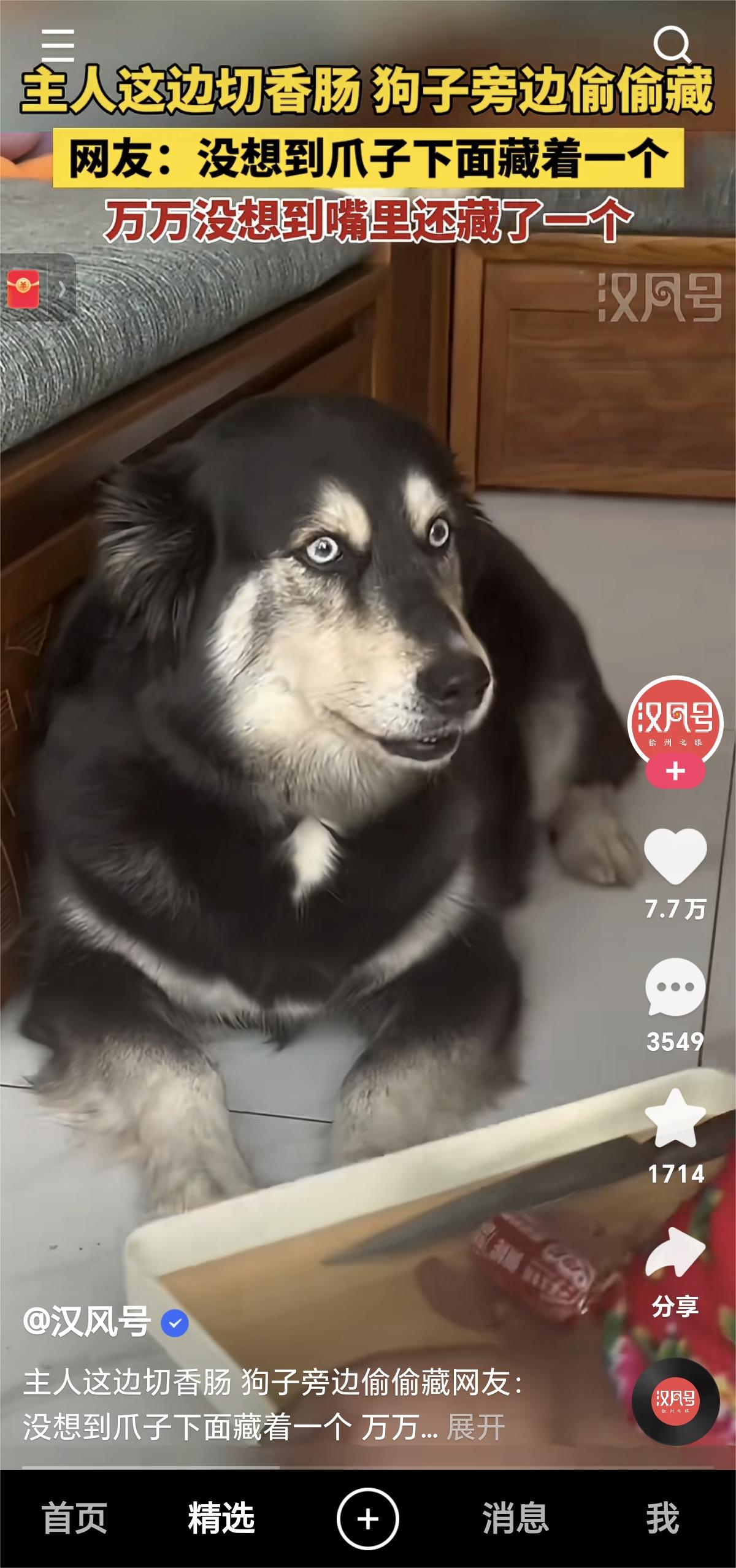}}
\subfigure[Examplar landing page of Bilibili from mobile app] {
\label{fig:example_ytb_app}
\includegraphics[width=.145\textwidth]{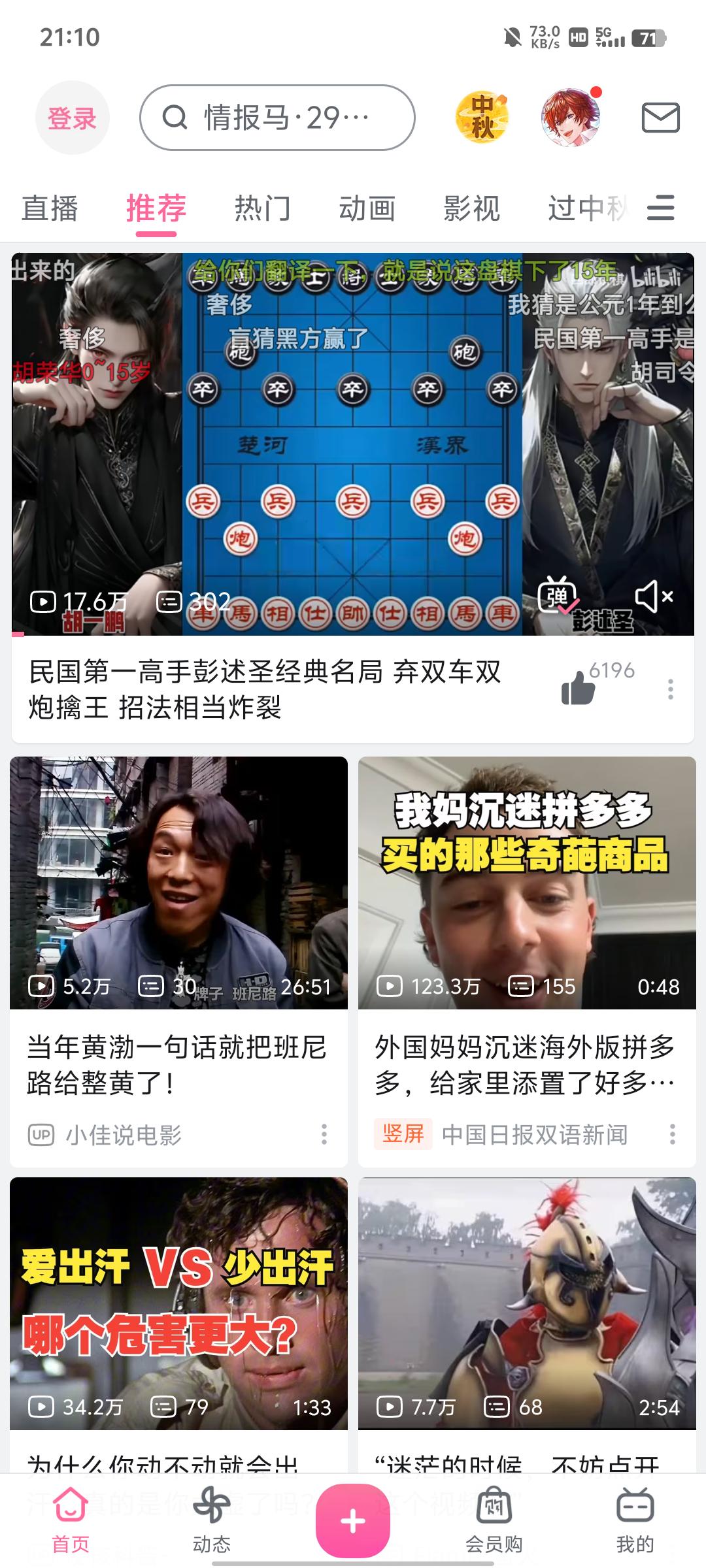}}
\subfigure[Examplar landing page of Bilibili from PC]
{\label{fig:example_ytb_pc}
\includegraphics[width=.36\textwidth]{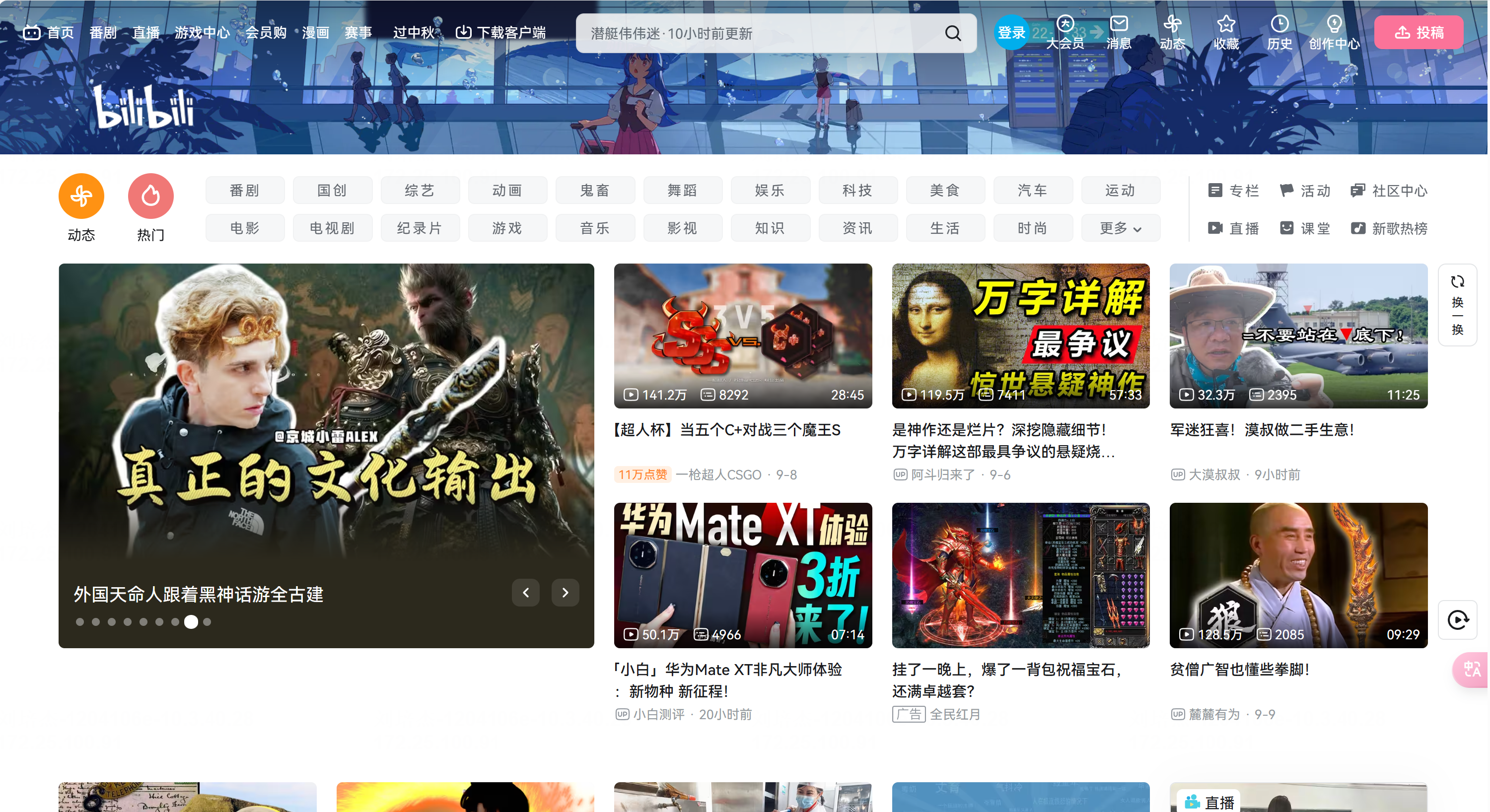}}
\caption{Examples of landing pages for Kuaishou and Bilibili.} \label{fig:example}
\end{figure}

Launched in 2011 and turned into a short video platform in 2012, Kuaishou has been one of the first pioneers in the short-form video industry in China~\cite{2020kuaishou}. It has grown to be one of the largest short-form video platforms in China, with more than 300 million daily average users as of 2023 and has been still enjoying fast growth. With ``Authentic, Diverse, Beautiful, and Beneficial'' as the core value and ``embrace all lifestyles'' as its motto, Kuaishou identifies itself as a content-based community welcoming vibrant engagement and social connections. Figure~\ref{fig:example} delineates a typical landing page of Kuaishou and its comparison with PC-based and app-based landing pages of Bilibili. Content creators upload videos with lengths ranging from seconds to minutes, and users land on the feed of algorithmically recommended videos upon opening the app (see Figure~\ref{fig:example_ks}), which is representative of the typical workflow of short-form video platforms. Switches between videos in the recommendation feed can be easily accomplished by simply scrolling down the screen on Kuaishou, and users no longer need to click through a video to play it or close it for exit (Figure~\ref{fig:example_ytb_app} and Figure~\ref{fig:example_ytb_pc}).

\subsection{Datasets}

\subsubsection{Kuaishou Dataset} 
We collect the metadata of all-round Kuaishou videos uploaded between July 29th and August 12th, 2024 to identify the characteristics of short-form videos. This dataset contains the basic information about the videos as well as users' aggregated interactions with the videos. Specifically, basic information about the videos records the publicly-attainable duration, upload date, anonymized and hashed author ID, and the category of the videos. Categorical information of the videos is arranged in a hierarchical manner, which is further aggregated into the following 7 major categories by 5 experts within the field: entertainment, film, gaming, knowledge, life, music, and outdoors. These are complemented with aggregated user interaction information detailing the number of views, likes, and shares that each video acquires within the first 7 days after they are uploaded, where the date of upload is treated as Day 1. Only videos viewed at least once in the first 7 days after their uploads are retained. In this way, information about 248.2 million short-form videos is gathered. 

\subsubsection{Bilibili Dataset} 
To enable direct comparison between short-form and long-form videos, we collect another dataset from Bilibili\footnote{https://www.bilibili.com/}, a leading long-form video-sharing platform in China. We choose this platform because (1) Bilibili represents the typical characteristics of long-form video-sharing platforms well, and (2) Kuaishou and Bilibili are both primarily based on Chinese audiences, which alleviates the possible confounding influence of cultural differences. Specifically, we crawl all videos uploaded to Bilibili between July 29th and August 12th, 2024. We limit our attention to the same time period to enhance the comparability of the two datasets. By first recording the links of the videos upon their uploads and then extracting user interaction with the videos after 7 days, we collect the basic information and aggregated user interaction records are provided. The basic information of a video includes its ID, title, uploader, time of release, duration, category, etc., and the aggregated user interaction records contain the number of views, likes, shares, collections, etc within the first 7 days since it is uploaded. Videos are classified into 16 categories, which are further grouped into the 7 major categories based on the expert knowledge of the same 5 experts: entertainment (dancing and entertainment), film (cartoon, film, and otomad), gaming, knowledge (news, knowledge, and technology), life (animals, autos, food, fashion, and life), music, and outdoors (sports). By retaining videos viewed at least once in the first 7 days since they are uploaded as Kuaishou, we arrive at 4.7 million videos.

\subsubsection{Ethical Considerations}
We treasure the rights of users and are fully aware of the sensitivity of the data. As such, careful measures are undertaken to address concerns over the usage and mining of the collected data. In terms of the Bilibili dataset, it is crawled from public data available to everyone. User-level details of interactions are aggregated, and only video-level characteristics are provided. In terms of the Kuaishou dataset, the data we collect is also publicly-attainable and carefully pre-processed. All the identifiable user-level information is aggregated, and only video-level characteristics are provided. Furthermore, consent for proper academic usage is included in Kuaishou's Terms of Service, and we have our research protocol reviewed and approved by our local institutional review board. Finally, the data is securely stored on an offline server, where only authorized individuals bounded by stringent non-disclosure agreements are able to access the data. As such, we make sure that users' privacy is properly protected.
\section{Basic Video Characteristics}

In this section, we investigate the basic characteristics of videos on Kuaishou and compare them with Bilibili videos. By looking into distributions concerning video lengths and video categories, we delineate what the videos are like (RQ1).

\subsection{Video length}

\begin{figure} [htb]
\centering
\subfigure[CDF of video length] {
\label{fig:duration_dis}
\includegraphics[width=.32\textwidth]{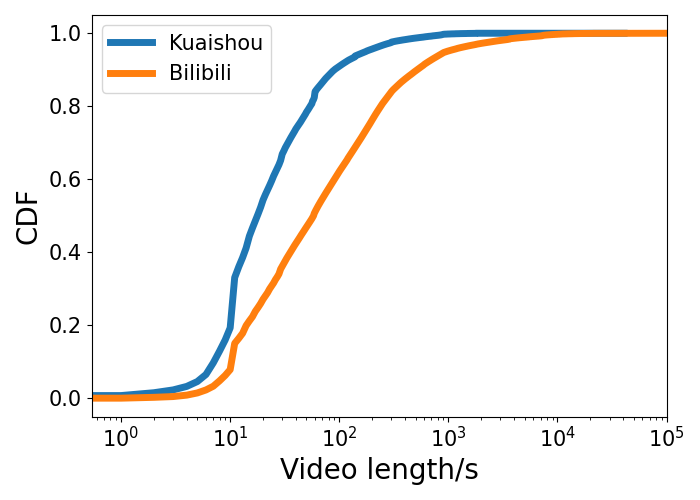}}
\subfigure[Video length distribution across various categories] {
\label{fig:cat_duration}
\includegraphics[width=.39\textwidth]{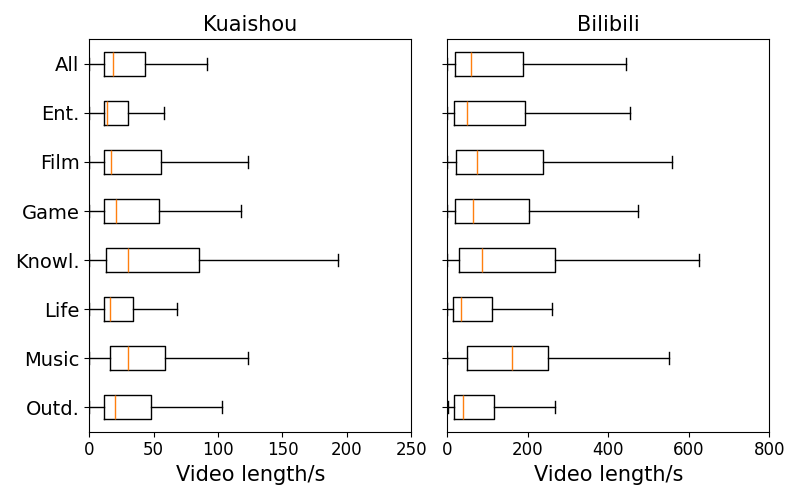}}

\caption{Comparison between video length distributions on Kuaishou and Bilibili.} \label{fig:duration}
\end{figure}

A frequently noted dimension of differences between short-form and long-form video platforms is the lengths of the videos. As such, we delve into the distribution of video lengths across Kuaishou and Bilibili. Figure~\ref{fig:duration_dis} shows the cumulative distribution functions (CDFs) of video lengths across the two platforms. Significant differences can be observed between the lengths of the videos on Kuaishou and Bilibili ($p<0.001$, Kolmogrov-Smirnov test), where the median length of Kuaishou videos is more than three times shorter than that of Bilibili (18s versus 59s). As indicated by the figure, 44.4\% of the videos on Kuaishou are no longer than 15 seconds. Videos in 1 minute constitute 83.0\% of all videos, and only 2.4\% are longer than 5 minutes. However, circumstances are vastly different for Bilibili, which predominantly consists of relatively longer videos. Only 21.2\% of the Bilibili videos are with lengths of at most 15 seconds. Videos within 1 minute account for a portion of 51.1\%, and 16.0\% of the videos on Bilibili share video lengths longer than 5 minutes.

These patterns remain consistent across different categories. In Figure~\ref{fig:cat_duration}, we show the distributions of the lengths of videos on Kuaishou and Bilibili, respectively. Although the lengths of the videos vary across categories, videos on Kuaishou are shorter than those on Bilibili by several times. For example, entertainment and life videos share median lengths of 14s and 16s on Kuaishou, whereas on Bilibili, these medians rise to 49s and 35s. Even for knowledge-related and music-related videos with relatively longer median lengths on Kuaishou (30s), they are still significantly shorter than their counterparts on Bilibili (median = 87s for knowledge and 161s for music).

\subsection{Category}

\begin{figure*} [tbp]
\centering
\subfigure[Number of videos] {
\label{fig:cat_num}
\includegraphics[width=.32\textwidth]{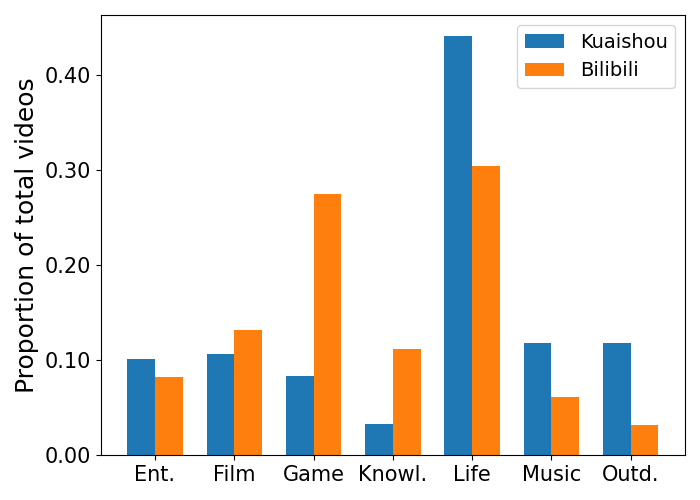}}
\subfigure[Number of views] {
\label{fig:cat_view}
\includegraphics[width=.32\textwidth]{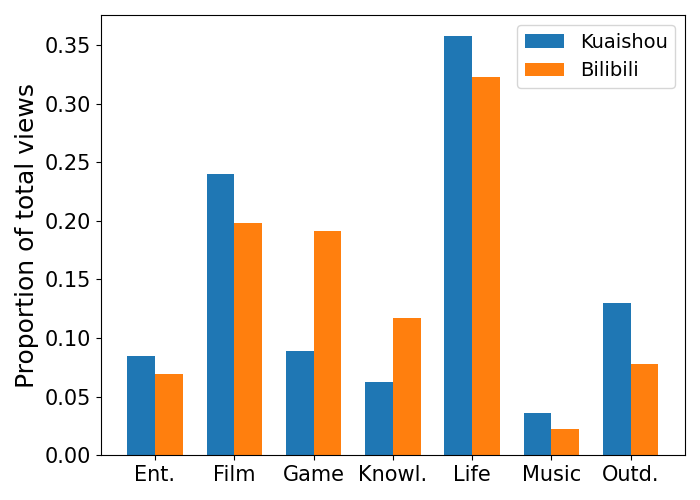}}
\subfigure[Relative ratio of the number of views and videos]
{\label{fig:cat_ratio}
\includegraphics[width=.32\textwidth]{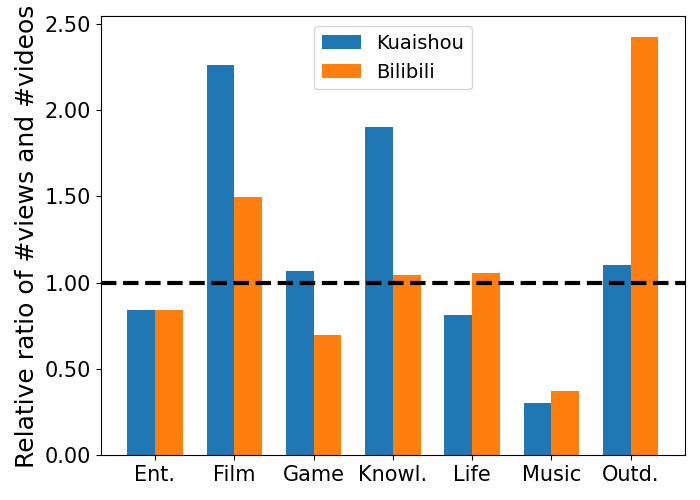}}
\caption{The number of videos, views, and their relative ratios across different categories on Kuaishou and Bilibili.} \label{fig:cat_dis}
\end{figure*}

The previous analysis indicates that across all categories, the lengths of videos shorten significantly on Kuaishou compared with Bilibili. However, do the production and consumption of different types of videos also vary? Figure~\ref{fig:cat_num} shows the proportion of video numbers for different categories on Kuaishou and Bilibili. Prominent discrepancies can be spotted in terms of the categorical distribution of videos across the two platforms ($\chi^2=3.29\times10^{6}$, $p<0.001$). Kuaishou is more over-represented in the categories of entertainment, life, music, and outdoors, while more under-represented in film, gaming, and knowledge than Bilibili. For example, videos pertaining to life account for the largest proportion of 44.0\%. Although life videos are also the most prevalent on Bilibili, the share it occupies drops to 30.5\%. In contrast, as the second most prevalent category on Bilibili, 27.5\% of Bilibili videos belong to gaming. However, only 8.3\% of Kuaishou videos are gaming-related. 

Similar discrepancies are also observed for the categorical distribution of the number of views ($\chi^2=4.19\times10^{9}$, $p<0.001$). As depicted in Figure~\ref{fig:cat_view}, videos concerning entertainment, film, life, music, and outdoors attract relatively more views as a whole compared with Bilibili, whereas the categories of gaming and knowledge show an opposite trend. For example, life videos, which receive the most views in all on Kuaishou, acquire 35.8\% of all views on Kuaishou. But this proportion is 32.2\% for Bilibili. However, for gaming videos, the corresponding proportions of total views for the two platforms are 8.8\% on Kuaishou and 19.1\% on Bilibili, respectively.

However, this distribution of views can be interrelated with the categorical distribution of videos. Taking this into account, we further investigate the average view counts for a video across different categories. As we show in Figure~\ref{fig:cat_ratio}, both similarities and discrepancies can be spotted between Kuaishou and Bilibili in terms of the average number of views per video. Both entertainment videos and music videos have relatively smaller ratios of view counts and video numbers, reflecting that videos pertaining to these categories are relatively less likely to be viewed per video across the two platforms. However, remarkable differences are also present. For example, the relative ratio of videos in the film and knowledge categories on Kuaishou are 2.24 and 1.72 respectively, both of which are far above 1, while the corresponding number for outdoor videos is 1.11. In comparison, for Bilibili videos, outdoor videos have the largest relative ratio of 2.42, followed by the category of film (1.50).  

Taken together, these results suggest that Kuaishou is relatively more over-represented by life-related videos, although the average view counts may be smaller. By contrast, the more interest-driven ones such as gaming and knowledge are more under-represented. Perhaps surprisingly, some informative categories such as knowledge seem to be relatively more likely to be viewed on average.

\section{User Engagement}
Apart from the characteristics of the videos themselves, how users interact and engage with these videos is also an integral concern of the ecology of video-sharing platforms~\cite{bartolome2023literature}. Therefore, in this section, we dig into user interaction and engagement patterns on Kuaishou and how it resembles and differentiates from Bilibili (RQ2). Taking the previously mentioned distinctive differences in video lengths into account, we focus especially on the interplay between user engagement and video lengths on Kuaishou and how it compares with Bilibili. Specifically, we explore the patterns concerning the relationships between video lengths and likes and sharing.

\subsection{Like}
\begin{figure} [htb]
\centering
\subfigure[Average number of likes for videos of different lengths] {
\label{fig:ave_like}
\includegraphics[width=.32\textwidth]{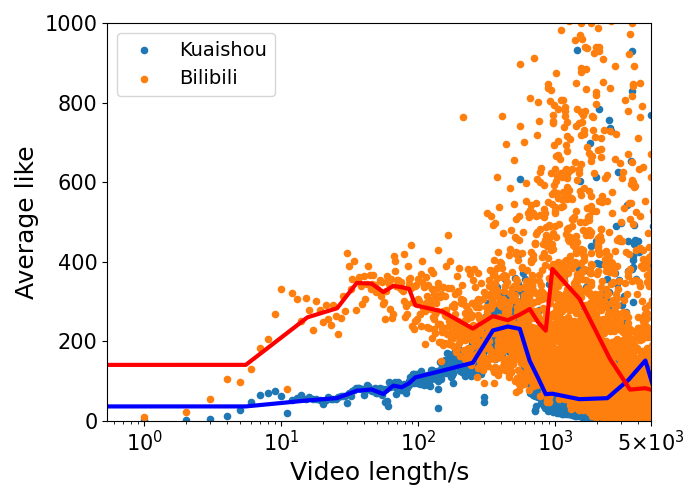}}
\subfigure[Average like rates for videos of different lengths] {
\label{fig:ave_like_rate}
\includegraphics[width=.32\textwidth]{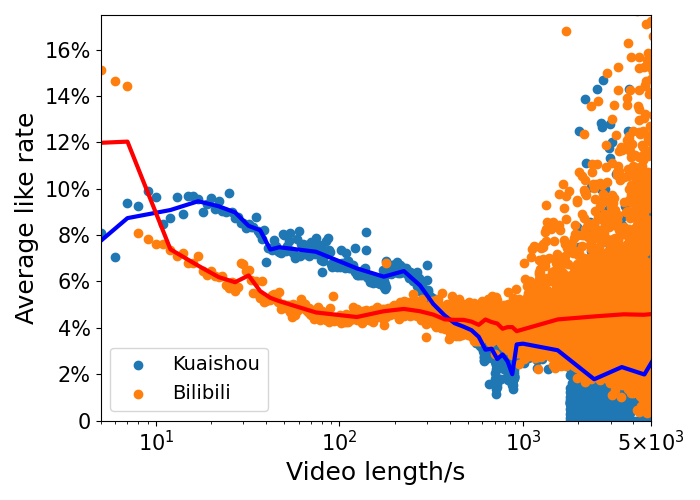}}
\caption{Average number of likes and like rates for videos of different lengths.} \label{fig:engagement_like}
\end{figure}

Likes have been important representations of user interactions and engagement on video-sharing platforms~\cite{wattenhofer2012youtube,munaro2021engage}. In Figure~\ref{fig:ave_like}, we show the numbers of likes a video receives across Kuaishou and Bilibili and how they change with respect to different video lengths. For Kuaishou, the average number of likes increases as the length of a video grows from 1 to 500 seconds. When videos become even longer, the average number of likes may decline. On Bilibili, a similar relationship is observed between the number of likes and video lengths on average, although when the length of the videos reaches 40s, the average number of likes remains relatively stable. However, an apparent negative relationship between time length and the average number of lengths can be observed when video length reaches 1000s. Moreover, the average number of likes seems to be larger for Kuaishou than for Bilibili, especially when the length of a video is relatively short.

Figure~\ref{fig:ave_like_rate} further describes the relationships between average like rates and video lengths across Kuaishou and Bilibili. Here like rates are calculated by dividing the number of likes by the number of views of a video to account for differences in audience size. Perhaps surprisingly, we discover that the average like rate of a video decreases as a video becomes longer on both Kuaishou and Bilibili. However, the drop on Kuaishou is a lot sharper than that on Bilibili: Kuaishou videos are liked at a rate of 9.3\%  at the length of 10s, which drops to 3.3\% at about 1000s; in comparison, the like rates of Bilibili videos declines from 6.7\% at 10s to 3.9\% at 1000s. These show that on a per-view basis, users show more preference for short videos, which is especially prominent on Kuaishou. Moreover, comparing the two lines, we find that the like rates are relatively larger for videos shorter than 400s, which characterizes the typical lengths of both Kuaishou and Bilibili. This indicates that videos are better at engaging users per view on Kuaishou. 


\subsection{Sharing}

\begin{figure} [htb]
\centering
\subfigure[Average number of shares for videos of different lengths] {
\label{fig:ave_share}
\includegraphics[width=.32\textwidth]{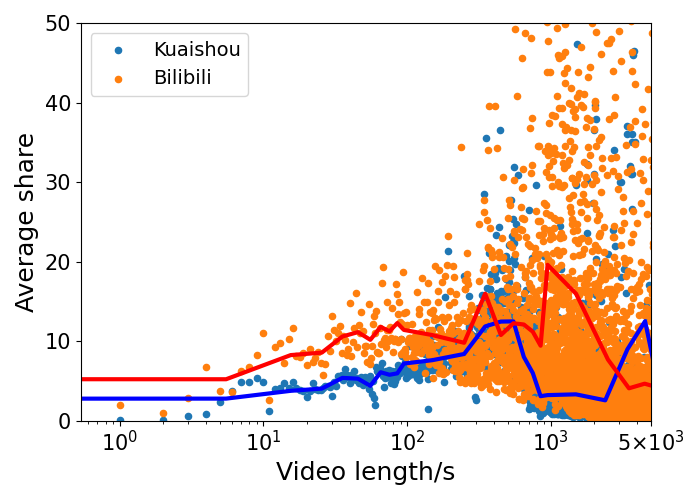}}
\subfigure[Average share rates for videos of different lengths] {
\label{fig:ave_share_rate}
\includegraphics[width=.32\textwidth]{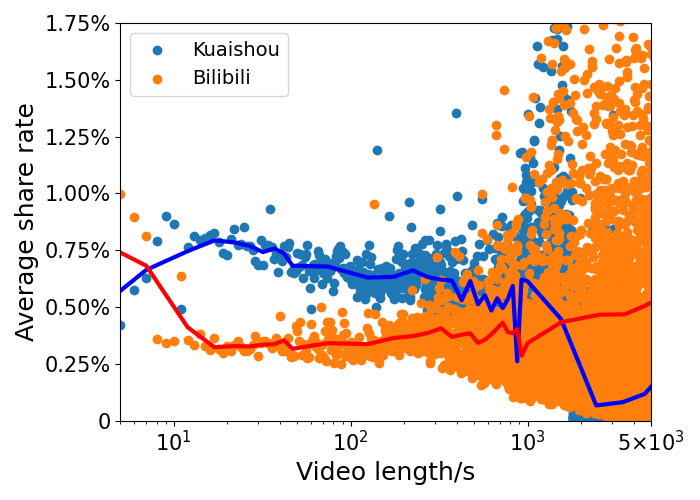}}
\caption{Average number of shares and average share rates for videos of different lengths.} \label{fig:engagement_share}
\end{figure}

Behaviors of sharing also signal important aspects of user interactions and engagement with a video~\cite{yang2022science}. Figure~\ref{fig:ave_share} displays the relationships between the average number of sharing a video receives in the first 7 days and its video length. On Kuaishou, the average sharing number gradually increases from 1 to 500 seconds, with a peak of about 12.5 at 500 seconds. When the length of a video further increases, the number of shares that it receives drops. In comparison, the average number of shares of a Bilibili video increases slowly when the length of a video grows from 10 to 1000 seconds. But when the length of a video exceeds 1000s, the average number of video shares begins to drop as a video lengthens. Comparing the curves of Kuaishou and Bilibili, we discover that the average number of shares of Kuaishou is smaller than that of Bilibili across a broad spectrum of video lengths. 

However, one may argue that the fewer shares a video receives are due to the variation in the number of audiences. To address this, we further examine the relationship between share rate and video length across Kuaishou and Bilibili. Here the share rate of a video is defined as the number of shares divided by the number of views of a video, which characterizes users' propensity to share a video upon watching. As depicted in Figure~\ref{fig:ave_share_rate}, the average share rates decrease as the lengths of Kuaishou videos increase; however, they remain relatively stable across Bilibili videos with varying lengths and even increase a bit as videos become longer. For example, the average share rate of Kuaishou videos drops from 0.67\% at 10 seconds to 0.39\% at 1000 seconds, while stabilizes at about 0.34\% on Bilibili. Comparing Kuaishou and Bilibili, we can observe that the average share rates are mostly larger on Kuaishou than on Bilibili across videos with typical video lengths of the two platforms. These results indicate that users are more likely to enact the behavior of sharing on Kuaishou than on Bilibili.

\section{Collective Attention Allocation}
\begin{figure*} [htb]
\centering

\subfigure[CCDF of the number of views] {
\label{fig:view_ccdf}
\includegraphics[width=.32\textwidth]{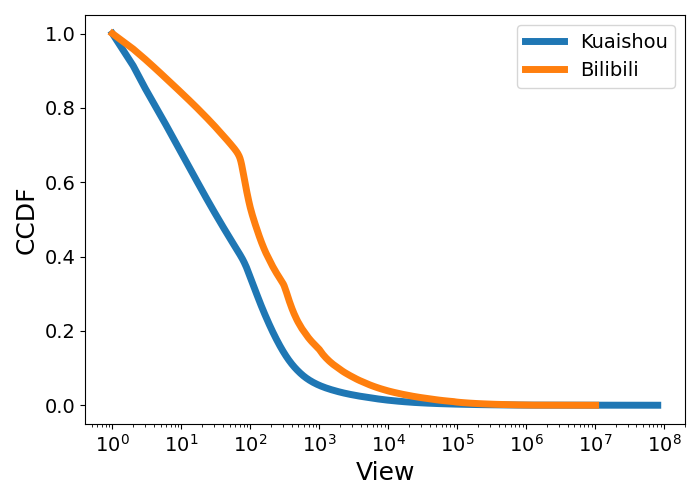}}
\subfigure[CDF of the proportion of views] {
\label{fig:view_portion_cdf}
\includegraphics[width=.32\textwidth]{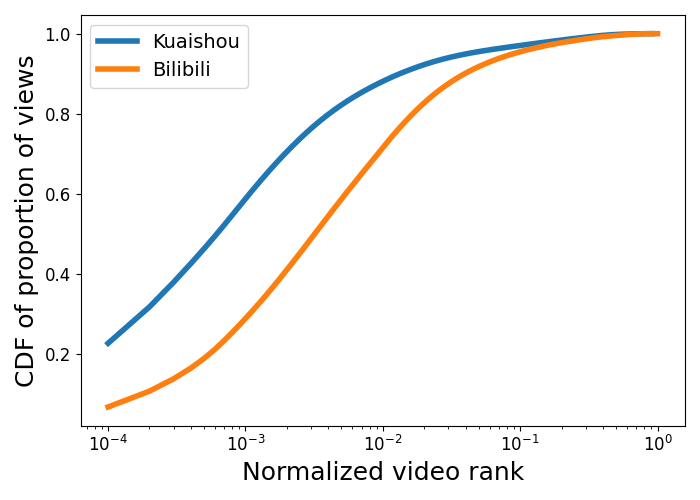}}
\caption{Distribution of views across videos.} \label{fig:dis_view}
\end{figure*}

The previous sections take into account the characteristics of a single video and its corresponding interactions with users. In this section, we take a more holistic and ecology-based view to consider video-to-video relationships and how users' collective attention is allocated among videos. Here we calculate a video's acquired attention by the number of views it acquires in the first 7 days after it is released.

\subsection{Video-level distribution}
One fundamental dimension of the allocation of users' collective attention is the per-video distribution of view counts. As shown in Figure~\ref{fig:view_ccdf}, the distributions of per-video view counts for Kuaishou and Bilibili videos are significantly different (Kolmogrov-Smirnov test, $p<0.001$). Compared with Bilibili, Kuaishou videos are more over-represented by the two ends, i.e., videos relatively less frequently viewed and extremely frequently viewed. 36.1\% of Kuaishou videos are viewed no more than 10 times, whereas the same proportion drops to 15.9\% for Bilibili.  However, only 38 (0.00080\%) Bilibili videos are viewed more than 5 million times, while about 7,000 (0.0026\%) Kuaishou videos acquire more than 5 million views. Further quantification shows that the Gini coefficient of view counts is larger on Kuaishou (0.977) than on Bilibili (0.959), which corroborates that the distribution of collective attention is more biased towards top videos on Kuaishou.

This unbalance in view counts has led to top video's dominance in the acquisition of users' collective attention as a whole. Figure~\ref{fig:view_portion_cdf} shows the distribution of the accumulation of views under different normalized rankings, where most frequently viewed videos are ranked with relatively smaller numbers. As can be seen from Figure~\ref{fig:view_portion_cdf}, 6.8\% of the views are attracted by the top 0.01\% most popular videos on Bilibili, and this proportion rises to 28.9\% for the top 0.1\% most popular videos. The last 80\% of the videos account for only 2.3\% of the views. However, the top 0.01\% of the popular videos account for 22.7\% of the views among Kuaishou videos with at least one view; the top 0.1\% most popular videos grab 58.8\% of all views, and the last 80\% of the videos only capture 1.6\% of all views on Kuaishou. As such, users' collective attention tends to be more dominated by the most frequently viewed videos on Kuaishou.

\subsection{Creator-level shift}

\begin{figure*} [htb]
\centering
\subfigure[The shift of the number of views for videos from the same creator] {
\label{fig:shift}
\includegraphics[width=.32\textwidth]{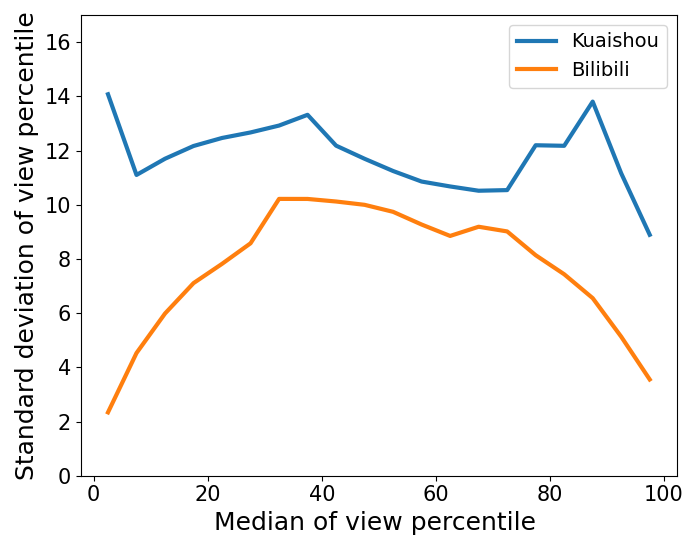}}
\subfigure[The probability of hit videos for creators] {
\label{fig:median_prob}
\includegraphics[width=.32\textwidth]{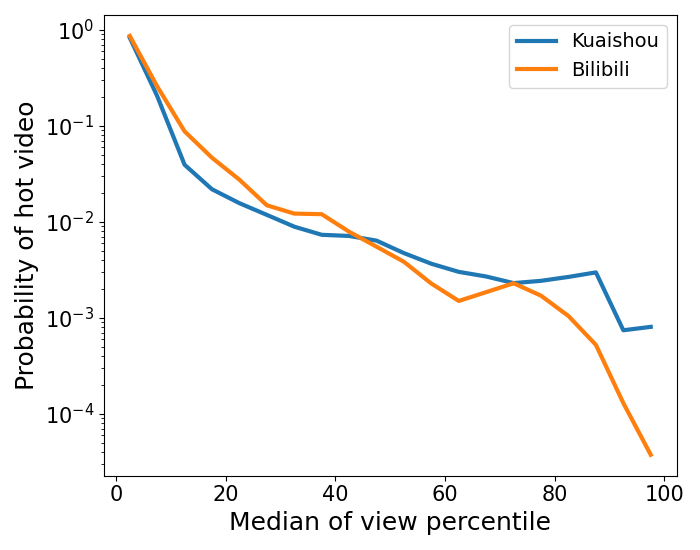}}
\subfigure[The relative ratio of the probability of hit videos between Kuaishou and Bilibili] {
\label{fig:ratio_prob}
\includegraphics[width=.32\textwidth]{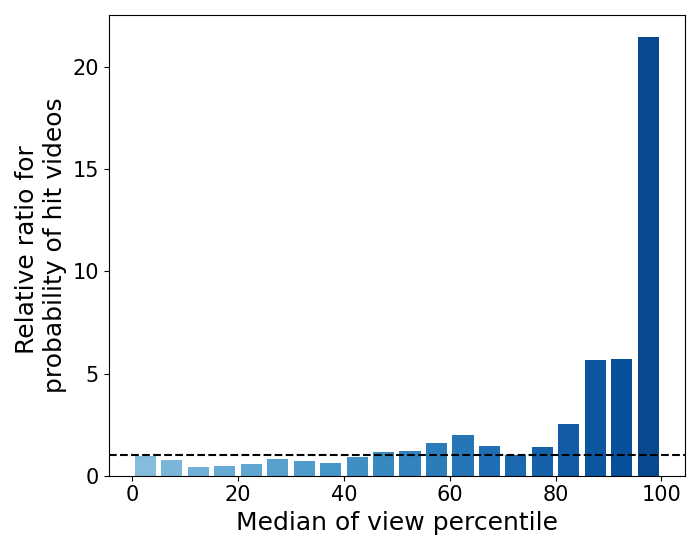}}
\caption{Creator-level shifts in popularity and the probability of posting hit videos across Kuaishou and Bilibili.} \label{fig:fairness}
\end{figure*}

So far we have indicated the relatively larger unevenness of videos on Kuaishou. However, this does not directly translate into the unbalance between creators because how videos released by the same content creator are like remains unclear. Therefore, in Figure~\ref{fig:shift}, we explore the relationship between the standard deviation and the median of the view percentile of a content creator's videos on Kuaishou and Bilibili, where more popular videos are ranked as smaller in percentile ranks. Here we retain only creators uploading no fewer than three videos to reduce the bias introduced by limited records. From the figure, we can observe that the standard deviation of the number of views from the same creator on Kuaishou remains relatively stable when their median number of views varies; but on Bilibili, an inverted U-shape is present. These 
demonstrate that the attention allocated to top and low-tier content creators on Bilibili remains relatively stable, with popular creators consistently attracting a large pool of audience and less popular ones consistently acquiring relatively few views. In contrast, attention levels on Kuaishou fluctuate significantly across the spectrum, showing remarkable ecological differences and possible creator-level changes. Comparing results from Kuaishou and Bilibili, we observe that the standard deviations of view percentiles of the same Kuaishou creators' videos are consistently higher than Bilibili videos across creators with different median view percentiles. This indicates that the rankings of view counts for videos uploaded by the same creators shift more variably on Kuaishou.

To understand what this variable shift brings to content creators, we measure the probability that a content creator posts a hit video across different median view percentiles (see Figure~\ref{fig:median_prob}). Here hit videos are identified as the top 5\% videos with the most number of views. Although the probability of hit videos drops significantly as the median view percentiles increase for both Kuaishou and Bilibili, it remains relatively higher for Kuaishou than for Bilibili when the median view percentiles are controlled. Indeed, as depicted in Figure~\ref{fig:ratio_prob}, the relative ratio of the probability of getting posted videos to be hit ones between Kuaishou and Bilibili is above 1 for authors with median view percentiles that are larger than 50\%, and is especially higher for creators with large median view percentiles (i.e., creators with relatively lower median video views). This indicates that ordinary content creators on Kuaishou are more likely to get posted videos hit, which is especially true for low-ranking content creators.

\section{Discussions}
By examining the video characteristics, user engagement, and collective attention allocation patterns of Kuaishou, our results shed light on the distinctiveness of short-form video platforms. Specifically, one prominent feature of short-form video platforms is the shortened video lengths. As our results demonstrate, the average duration of videos on short-form platforms is several times shorter than those on long-form video-sharing platforms. This change in video length may alter the types of content suitable for dissemination, which, in turn, could lead to shifts in the categorical distribution of video production and consumption. The results of this can be at least two-fold. Firstly, similar to the distinction between UGC and non-UGC platforms where the shortened lengths reduce the required efforts for production~\cite{cha2007tube}, short-form video platforms further lower the barrier for production by further reducing video lengths. This is likely to explain the relatively more life-related short-form videos observed on these platforms. Secondly, content creators, especially those migrating from traditional long-form video-sharing platforms, need to adopt special strategies to align with the ecosystem of short-form video platforms. For example, they may make platform-specific content curations, such as distillation and segmentation~\cite{ma2023multi}. 

Our results also indicate important dimensions regarding user engagement with short-form videos. First, users explicitly engage with the videos less per view on the short-form video platform across the dimensions of likes and sharing. This may be surprising considering the different video play modes between short-form and long-form videos. On a typical long-form video platform, a user needs to click a video before getting the video started to play. This action of clicking filters out a large proportion of videos that seem uninteresting to the user. However, on short-form video platforms, this action of clicking through is removed, which means that all videos appearing in the recommendation feeds will be directly played for a user. In this way, videos that the user is uninterested in are also counted as played for the user, and the rates of likes and sharing should have been lower for short-form video platforms than long-form video-sharing platforms. It is therefore surprising to find that considering these seemingly disadvantages, short-form videos still share relatively higher like rates and share rates. The effective modeling of user interests enabled by recommender systems should contribute to this~\cite{barta2021constructing,karizat2021algorithmic}, and we leave it for future work to unpack the nuanced and exciting mechanisms underlying the higher engagement per view for short-form videos. Second, we find that the relationship between video lengths and user engagement can vary substantially when the indicator used for measuring user engagement changes. For example, video length positively correlates with the average number of likes and sharing, but is negatively associated with average like rates and sharing rates, which is especially prominent on Kuaishou. This highlights that future scholars and practitioners should be cautious about the targeted indicator for dissecting user engagement, and take into account whether the number of counts or rates per view should be chased for.


Moreover, by delving into the allocation of collective attention over videos, we find that view counts are biased towards top videos on the short-form video platform. With relatively fewer sharing behaviors and rates delineating the relatively fewer social behaviors, this further highlights the essential mediation of the underlying recommendation algorithm. Indeed, 90\% of the consumed videos are from algorithmic recommendations on Kuaishou. This rich-get-richer phenomenon is somewhat expected under an algorithm-mediated circumstance, because user-level interest in a video is clearer and easier to capture for recommender systems when it interacts with more users. To alleviate this, platform practitioners are encouraged to design better approaches to addressing cold start modeling of video features and figure out better ways to empower long-tail videos. However, this does not translate into harm for ordinary content creators as a whole. The video count percentiles for videos uploaded by the same content creator fluctuate more, and non-top content creators have relatively higher probabilities of producing hit videos. This means that the mediated recommendation system creates an even fairer ecology on the creator level, providing more chances for ordinary content creators to stand out. This should be especially praised of, and we encourage future research to identify how this is achieved and how this could be learned by other platforms. 

\subsection{Limitations, future work, and potential broader impact}
Firstly, our delineation of the short-form video platform comes primarily from one single platform, i.e., Kuaishou. Therefore, there may be possibilities of context bias for our findings. Future work is encouraged to extend our current work to multiple short-form videos to distinguish platform-specific designs and the characteristics of short-form video ecology for a better understanding of the nature of short-form video platforms. Secondly, Kuaishou is based on the Chinese context. Differences in culture and audience base may play a role in how users treat and interact with the videos on the platform. Although we try to eliminate the confounding influence of culture by comparing it with a leading long-form video-sharing platform, Biilibili, future studies can proceed towards short-form videos in other cultural backgrounds to verify the generalizability of the current discoveries. However, Kuaishou has been among one of the earliest and the most prevalent short-form video platforms, which makes the case of Kuaishou itself worthy of investigation. The workflow on Kuaishou is representative of the typical workflow of short-form video platforms in general, and the large-scale all-round investigation of Kuaishou videos also eliminates the possibilities of potential selective bias. Therefore, we believe the current results should point to intrinsic characteristics of short-form videos, which have to potential to apply to a wide range of specific circumstances of short-form video platforms.

\section{Conclusion}
In this paper, we show how short-form video platforms deviate from traditional video-sharing platforms through conducting large-scale data-driven investigations of Kuaishou videos and comparing them with Bilibili. We find the videos are several times shorter on average, which share distinctively different categorical distributions. Users explicitly interact and engage with the videos less per view on Kuaishou across dimensions of engagement such as likes and sharing. Collective attention is more biased towards top videos. Moreover, videos uploaded by the same content creators vary more significantly in view counts, where ordinary creators have relatively higher probability to popularize their videos. Taken together, our work unravels the distinctiveness of the emerging short-form videos, which lies the bases for future related research and design of the increasingly popular short-form video ecology. 

\bibliography{aaai22}

\bigskip

\end{document}